\documentclass[aps,pra,twocolumn]{revtex4-1}

\usepackage{latexsym}
\usepackage{float}

\usepackage{mathrsfs}
\usepackage{graphicx,amsfonts,amssymb,amsmath,braket,hyperref}
\usepackage{epstopdf}

\begin{document}


\title{Spin-Orbit Coupling Induced Resonance in an Ultracold Bose Gas}



\author{Qi Gu, Lan Yin}
\email[]{yinlan@pku.edu.cn}


\affiliation{School of Physics, Peking University, Beijing 100871, China}


\date{\today}

\begin{abstract}
We study a two-component Bose and gas with artificial spin-orbit coupling (SOC) which couples the center-of-mass momentum of atom to its internal states.  We show that in this system resonance can be induced by tuning SOC strength.  With a two-dimensional SOC, resonances in two scattering channels can be induced by tuning the aspect ratio of SOC strengths.  With a three-dimensional
SOC, resonance in all three scattering channels can be induced by tuning the appropriate SOC strength.  Our method can also be applied to a Fermi gas where resonance can also be induced with two- or three-dimensional SOC.
\end{abstract}



\maketitle

\textit{Introduction}.
In ultracold quantum gases, resonance scattering can often be induced by various means. In alkali atoms, the effective interaction between atoms can be tuned by magnetic Feshbach resonance \cite{MFR} in which scattering states in the open channel are coupled to molecular states in the closed channel and their energy difference can be tuned by an external magnetic field.  In alkali earth atoms, the effective interaction can be tuned by optical Feshbach resonance \cite{OFR} in which a laser field couples two scattering atoms in the open channel to a molecular state in the close channel.  In trapped system, confinement induced resonance \cite{CIR} may occur, which can be also interpreted as due to coupling between effective closed and open channels similar to a Feshbach resonance \cite{CIR2}.  These methods of inducing resonance are powerful tools to manipulate ultracold quantum gases.

Artificial spin-orbit coupling (SOC) which couples the center-of-mass momentum and internal states of an atom \cite{REV1,REV2,REV3,REV4}, was realized in experiments on Bose gases \cite{BSOC1,BSOC2} and Fermi gases \cite{FSOC1,FSOC2}.  Although so far most of experimental SOC was one-dimensional (1D), recently two-dimensional (2D) SOCs were successfully generated in experiments \cite{2SOC1,2SOC2}.  In a Fermi gas with a 1D SOC, it was found experimentally that near a Feshbach resonance the resonance position can be shifted by changing the detuning energy and intensity of Raman lasers \cite{socir1}, which was also studied theoretically \cite{socir2,socir3}.  In a two-component Bose gas with SOC, the low-energy scattering problem is even more complicated because there are three s-wave scattering channels instead of one.  In this work, we theoretically investigate how to induce resonances in a Bose gas with SOC.

We will show that in a two-component Bose gas with a general three-dimensional (3D) anisotropic SOC resonances can be induced by tuning SOC strengths, as a result of the special single-particle excitation and low-energy density of states (DOS) due to SOC. Our main results are as follows.  In a Bose gas with a 2D anisotropic SOC, resonance in two scattering channels can be induced by tuning the aspect ratio of SOC strengths, while the other scattering channel is unaffected.  In a Bose gas with a 3D anisotropic SOC, resonance in all three scattering channels can be induced.  The resonance position in each scattering channel can be tuned more effectively by changing the SOC strength in the corresponding direction, which can be very useful for studying spin-dependent effects.  In the same formulism, we study a Fermi gas with a 2D or 3D SOC  where resonance can also be induced by tuning SOC strength in any direction.  The implication of our results to current experiments is also discussed.

\textit{Model}.
We first consider a two-component homogeneous Bose gas with a SOC, described by the Hamiltonian $H=H_0+H_\text{soc}+H_\text{int}$.  The SOC term is given by
\begin{equation}
H_\text{soc}=\sum_{\mathbf{k}\rho\rho'}c_{\mathbf{k}\rho}^{\dagger} \mathbf{h}_{\mathbf{k}}^{\phantom\dagger} \cdot {\boldsymbol \sigma}_{\rho\rho'}^{\phantom\dagger} c_{\mathbf{k}\rho'}^{\phantom\dagger},
\end{equation}
where $c_{\mathbf{k}\rho}$ is the annihilation operator of a boson with wavevector $\mathbf{k}$ and spin component $\rho$, $\rho=\uparrow$ or $\downarrow$, $h_{\mathbf{k}\alpha}=\hbar^2 \lambda_\alpha k_\alpha/m$, $m$ is the mass of an atom, $\lambda_\alpha$ is the strength of SOC in $\alpha$-direction, $\alpha=x,y,z$, and $\sigma_\alpha$ is the Pauli matrix, $\sigma_{x\rho\rho'}=1-\delta_{\rho\rho'}$, $\sigma_{y\rho\rho'}=-i\delta_{\rho\uparrow}\delta_{\rho'\downarrow}+i\delta_{\rho\downarrow}\delta_{\rho'\uparrow}$, $\sigma_{z\rho\rho'}=\delta_{\rho\rho'}(\delta_{\rho\uparrow}-\delta_{\rho\downarrow})$,   The SOC becomes Rashba SOC when $\lambda_x=\lambda_y$ and $\lambda_z=0$, and Weyl SOC when $\lambda_x=\lambda_y=\lambda_z$.

\begin{figure}[h!]
		\centering
		\includegraphics[width=0.3\textwidth]{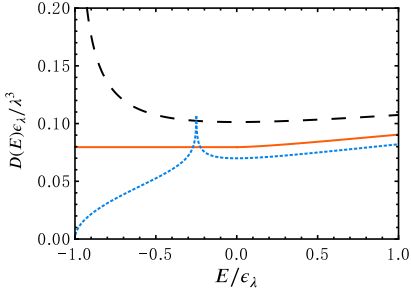}
		\caption{\label{DOS}DOS of a boson with SOC.  The solid line is the DOS with Rashba SOC which is a step function at energy minimum.  The dashed line is the DOS with Weyl SOC which is inversely proportional to the square root $\sqrt{E+\epsilon_\lambda}$ near energy minimum.  The dotted line is DOS with SOC $\lambda_z=2\lambda_x=2\lambda_y$ which is proportional to the square root $\sqrt{E+\epsilon_\lambda}$ near energy minimum, similar to the case without SOC.}
\end{figure}

The kinetic energy term is given by $ H_0=\sum_{\mathbf{k}\rho}c_{\mathbf{k}\rho}^{\dagger} \epsilon_\mathbf{k} c_{\mathbf{k}\rho}^{\phantom\dagger} $ where $\epsilon_\mathbf{k}=\hbar^2 k^2/2m$.  The single-atom Hamiltonian $H_0+H_\text{soc}$ can be diagonalized, yielding two helical excitation branches $\varepsilon_{\mathbf{k}}^{\pm}=\epsilon_\mathbf{k}\pm h_\mathbf{k}$ where $h_\mathbf{k}=|\mathbf{h}_\mathbf{k}|$.  The energy minimum of the lower branch $\varepsilon_{\mathbf{k}}^-$ is given by $-\epsilon_\lambda$ where $\lambda=\max(|\lambda_x|,|\lambda_y|,|\lambda_z|)$.  In the case of Rashba and Weyl SOCs, the density of states (DOS) of the lower branch, $D(E)=\sum_\mathbf{k} \delta(E-\varepsilon_{\mathbf{k}}^-)/V$ where $V$ is volume, are qualitatively different near the energy minimum from that without SOC, as shown in Fig.~\ref{DOS}.  The low-energy DOS has a strong effect on bound-state energies and resonance positions as discussed later in this paper.

We consider low-energy effective interactions between bosons, which include three contact interactions, two intraspecies and one interspecies, given by
	\begin{equation}
		H_{\text{int}}=\frac{1}{2V}\sum_{\mathbf{k}\mathbf{k'}\mathbf{q}\rho\rho'}  g_{\rho\rho'}^{\phantom\dagger}c_{\mathbf{k}^\prime+\mathbf{q}\rho}^\dagger c_{-\mathbf{k}^\prime+\mathbf{q}\rho'}^\dagger c_{-\mathbf{k}+\mathbf{q}\rho'}^{\phantom\dagger} c_{\mathbf{k}+\mathbf{q}\rho}^{\phantom\dagger}.
	\end{equation}
Due to symmetry we have $g_{\uparrow\!\downarrow}=g_{\downarrow\!\uparrow}=4 \pi \hbar^2 a'/m$, where $a'$ is the interspecies scattering length in the absence of SOC.  In the following, for simplicity, we consider the case with the same intraspecies interaction, $g_{\uparrow\!\uparrow}=g_{\downarrow\!\downarrow}=4 \pi \hbar^2a/m$, where  $a$ is the intraspecies scattering length in the absence of SOC.

\textit{Two-body bound states}.
The eigenequation of a two-body bound state is given by  $H\ket{\varphi} = E_0 \ket{\varphi}$, where $E_0$ and $\ket{\varphi}$ are eigenenergy and eigenstate of a bound state with zero center-of-mass momentum,
	\begin{equation}
		 \ket{\varphi}=\frac{1}{2}\sum_{\mathbf{k}\rho\rho^\prime}\phi_{\rho\rho^\prime}(\mathbf{k},-\mathbf{k})c_{\mathbf{k}\rho}^\dagger c_{-\mathbf{k}\rho^\prime}^{\dagger}\ket{0}.
	\end{equation}
To obtain the coefficient $\phi_{\rho\rho^\prime}(\mathbf{k},-\mathbf{k})$ and eigenenergy $E_0$, we follow the approach in Ref.~\cite{BCSOC2,BCSOC3} and rewrite the eigenequation as
	\begin{equation}\label{eigeneq}
		M_\mathbf{k} \Phi_\mathbf{k}=\frac{1}{V} G \sum_\mathbf{p}\Phi_\mathbf{p}  ,
	\end{equation}
where $\Phi_\mathbf{k}$ is a newly-defined vector with four components, $\Phi_\mathbf{k}=[ \phi_{\uparrow\!\uparrow}(\mathbf{k},-\mathbf{k}), \phi_{\downarrow\!\downarrow}(\mathbf{k},-\mathbf{k}), \phi_{\uparrow\!\downarrow}(\mathbf{k},-\mathbf{k}), \phi_{\downarrow\!\uparrow}(\mathbf{k},-\mathbf{k})]^T$, and $G$ is the matrix of coupling constants
	\begin{equation}
		G=
			\begin{bmatrix}
				g_{\uparrow\!\uparrow} & {\phantom{0}}0{\phantom{0}} & 0 & 0 \\
				{\phantom{0}}0{\phantom{0}} & g_{\downarrow\!\downarrow} & 0 & 0 \\
				{\phantom{0}}0{\phantom{0}} & {\phantom{0}}0{\phantom{0}} & g_{\uparrow\!\downarrow} & 0 \\
				{\phantom{0}}0{\phantom{0}} & {\phantom{0}}0{\phantom{0}} & 0 & g_{\downarrow\!\uparrow} \\
			\end{bmatrix}.
	\end{equation}
The matrix $M_\mathbf{k}$ is given by
	\begin{equation}
		M_\mathbf{k}=
			\begin{bmatrix}
				\xi_\mathbf{k} & 0 & S^*(\mathbf{k}_\bot) & -S^*(\mathbf{k}_\bot) \\
				0 & \xi_\mathbf{k} & -S(\mathbf{k}_\bot) & S(\mathbf{k}_\bot) \\
				S(\mathbf{k}_\bot) & -S^*(\mathbf{k}_\bot) & \xi_\mathbf{k}-2h_{kz} & 0 \\
				-S(\mathbf{k}_\bot) & S^*(\mathbf{k}_\bot) & 0 & \xi_\mathbf{k}+2h_{kz} \\
			\end{bmatrix} ,
	\end{equation}
where $\xi_\mathbf{k}=E_0-2\epsilon_{\mathbf{k}}$ and $S(\mathbf{k}_\bot)=h_{\mathbf{k}x} + ih_{\mathbf{k}y}$.
Define $Q=G \sum_\mathbf{k} \Phi_\mathbf{k}/V$, and following Eq.~\eqref{eigeneq} we obtain
	\begin{equation}\label{Qeq}
		Q=\frac{1}{V} G \sum_\mathbf{k} M_{\mathbf{k}}^{-1} Q.
	\end{equation}

In general Eq.~\eqref{Qeq} has four solutions.  We find that there are three physical bound states, two intra-species bound states with $Q_1=[ q_1,-q_1,0,0 ]$ and $Q_2=[ q_2,q_2,0,0 ]$, and one inter-species bound state with $Q_3=[0,0,q_3,q_3]$.  The other solution, $Q_4=[0,0,q_4,-q_4]$, satisfies Fermi statistics instead of Bose statistics and is thus ignored.  Further more, we find that although the eigenstate of a single atom mixed up two spin components, eigenenergies of the three two-body bound-states are determined by either intraspecies scattering length $a$ or interspecies scattering length $a'$, not both.  From Eq.~\eqref{Qeq} (see Appendix A for the derivation), we obtain equations for eigenenergies of intra-species bound states
\begin{equation}\label{intra-}
		\frac{m}{4\pi \hbar^2a}=\frac{1}{V}\sum_\mathbf{k}\left[\frac{1}{2\epsilon_\mathbf{k}} + \frac{\xi_{\mathbf{k}}^2-4h_\mathbf{k}^2 + 4h_{\mathbf{k}\alpha'}^2}{\xi_{\mathbf{k}}(\xi_{\mathbf{k}}^2-4h_\mathbf{k}^2)} \right],
\end{equation}
and of the inter-species bound state
	\begin{equation}\label{inter-}
		\frac{m}{4\pi \hbar^2a^\prime}=\frac{1}{V}\sum_\mathbf{k}\left[\frac{1}{2\epsilon_\mathbf{k}} + \frac{\xi_{\mathbf{k}}^2-4h_\mathbf{k}^2 + 4h_{\mathbf{k}z}^2}{\xi_{\mathbf{k}}(\xi_{\mathbf{k}}^2-4h_\mathbf{k}^2)} \right],
	\end{equation}
where $\alpha'=x$ for bound state $Q_1$ and $\alpha'=y$ for bound state $Q_2$.  In Eq.~\eqref{intra-} and~\eqref{inter-}, if $\lambda_\alpha=0$, $h_{\mathbf{k}\alpha}=0$, then the eigenenergy equation of the corresponding bound state is the same as that without SOC.  When $\lambda_\alpha \neq 0$, the eigenenergy of corresponding bound state clearly has SOC dependence.  Especially in the cases of Rashba and Weyl SOCs, Eq.~\eqref{intra-} and~\eqref{inter-} have infrared divergences at energy threshold $E_0=-2\epsilon_\lambda$ on r.h.s, due to the special low-energy DOS.  Resonance occurs at the scattering length $a_\text{r}$ where the binding energy of the bound state $E_b=-E_0-2\epsilon_{\lambda}$ vanishes.  After eigenenergies are solved, the bound state wave functions can be easily obtained from the Eq. \eqref{eigeneq}, as shown in Table \ref{table}.  Bound states $Q_1$ and $Q_3$ satisfy
$\phi_{\uparrow\!\uparrow}(\mathbf{k},-\mathbf{k})=-\phi^*_{\downarrow\!\downarrow}(\mathbf{k},-\mathbf{k})$, while the bound state $Q_2$ satisfies $\phi_{\uparrow\!\uparrow}(\mathbf{k},-\mathbf{k})=\phi^*_{\downarrow\!\downarrow}(\mathbf{k},-\mathbf{k})$.  All $Q_1$, $Q_2$ and $Q_3$ bound states satisfy the symmetry $\phi_{\rho\!\rho'}(\mathbf{k},\mathbf{k}')=\phi_{\rho'\!\rho}(\mathbf{k}',\mathbf{k})$.
\begin{table*}
	\caption{\label{table}Two-body bound state wavefunctions. The first column describes the bound-state type. The next four columns are coefficients of bound-state wavefunctions. Here $q_1,q_2,q_3$ and $q_4$ are normalization constants.}
	\begin{ruledtabular}
		\begin{tabular}{ccccc}
			Bound States  &  $\phi_{\uparrow\!\uparrow}(\mathbf{k},-\mathbf{k})$  &  $\phi_{\downarrow\!\downarrow}(\mathbf{k},-\mathbf{k})$  &  $\phi_{\uparrow\!\downarrow}(\mathbf{k},-\mathbf{k})$  &  $\phi_{\downarrow\!\uparrow}(\mathbf{k},-\mathbf{k})$ \\ \hline
			$Q_1$  &  $\frac{\xi_{\mathbf{k}}^2-4(h_{\mathbf{k}y}^2 + h_{\mathbf{k}z}^2+i h_{\mathbf{k}x} h_{\mathbf{k}y})}{\xi_\mathbf{k} \left(\xi_{\mathbf{k}}^2-4 h_\mathbf{k}^2\right)} q_1$  &  $\frac{4(h_{\mathbf{k}y}^2 + h_{\mathbf{k}z}^2-i h_{\mathbf{k}x} h_{\mathbf{k}y} )-\xi_{\mathbf{k}}^2}{\xi_\mathbf{k} \left(\xi_{\mathbf{k}}^2-4 h_\mathbf{k}^2\right)} q_1$   &  $\frac{-2h_{\mathbf{k}x}(\xi_{\mathbf{k}}+2h_{\mathbf{k}z})}{\xi_\mathbf{k} \left(\xi_{\mathbf{k}}^2-4 h_\mathbf{k}^2\right)} q_1$  &  $\frac{2h_{\mathbf{k}x}(\xi_{\mathbf{k}}-2h_{\mathbf{k}z})}{\xi_\mathbf{k} \left(\xi_{\mathbf{k}}^2-4 h_\mathbf{k}^2\right)} q_1$  \\
			$Q_2$  &  $\frac{\xi_{\mathbf{k}}^2-4(h_{\mathbf{k}x}^2 + h_{\mathbf{k}z}^2-i h_{\mathbf{k}x} h_{\mathbf{k}y})}{\xi_\mathbf{k} \left(\xi_{\mathbf{k}}^2-4 h_\mathbf{k}^2\right)} q_2$
			  &  $\frac{\xi_{\mathbf{k}}^2-4(h_{\mathbf{k}x}^2 + h_{\mathbf{k}z}^2+i h_{\mathbf{k}x} h_{\mathbf{k}y})}{\xi_\mathbf{k} \left(\xi_{\mathbf{k}}^2-4 h_\mathbf{k}^2\right)} q_2$  &  $\frac{-2i h_{\mathbf{k}y}(\xi_{\mathbf{k}}+2h_{\mathbf{k}z})}{\xi_\mathbf{k} \left(\xi_{\mathbf{k}}^2-4 h_\mathbf{k}^2\right)} q_2$  &  $\frac{2i h_{\mathbf{k}y}(\xi_{\mathbf{k}}-2h_{\mathbf{k}z})}{\xi_\mathbf{k} \left(\xi_{\mathbf{k}}^2-4 h_\mathbf{k}^2\right)} q_2$  \\
			$Q_3$  &  $\frac{-4h_{\mathbf{k}z}(h_{\mathbf{k}x}-i h_{\mathbf{k}y})}{\xi_\mathbf{k} \left(\xi_{\mathbf{k}}^2-4 h_\mathbf{k}^2\right)} q_3$  &  $\frac{4h_{\mathbf{k}z}(h_{\mathbf{k}x}+i h_{\mathbf{k}y})}{\xi_\mathbf{k} \left(\xi_{\mathbf{k}}^2-4 h_\mathbf{k}^2\right)} q_3$
			  &  $\frac{\xi_{\mathbf{k}}(\xi_{\mathbf{k}}+2h_{\mathbf{k}z})-4(h_{\mathbf{k}x}^2+h_{\mathbf{k}y}^2)}{\xi_\mathbf{k} \left(\xi_{\mathbf{k}}^2-4 h_\mathbf{k}^2\right)} q_3$  &  $\frac{\xi_{\mathbf{k}}(\xi_{\mathbf{k}}-2h_{\mathbf{k}z})-4(h_{\mathbf{k}x}^2+h_{\mathbf{k}y}^2)}{\xi_\mathbf{k} \left(\xi_{\mathbf{k}}^2-4 h_\mathbf{k}^2\right)} q_3$  \\
			$Q_4$  &  $\frac{-2(h_{\mathbf{k}x}-i h_{\mathbf{k}y})}{\xi_{\mathbf{k}}^2-4h_\mathbf{k}^2} q_4$  &  $\frac{2(h_{\mathbf{k}x}+i h_{\mathbf{k}y})}{\xi_{\mathbf{k}}^2-4h_\mathbf{k}^2} q_4$  &  $\frac{\xi_{\mathbf{k}}+2h_{\mathbf{k}z}}{\xi_{\mathbf{k}}^2-4h_\mathbf{k}^2} q_4$  &  $\frac{-\xi_{\mathbf{k}}+2h_{\mathbf{k}z}}{\xi_{\mathbf{k}}^2-4h_\mathbf{k}^2} q_4$
		\end{tabular}
	\end{ruledtabular}
\end{table*}

\textit{2D anisotropic SOC}.
We first consider 2D SOC with $\lambda_z=0$.  In the case of Rashba SOC with $\lambda_x=\lambda_y$, due to the non-vanishing DOS, the resonance position of two intraspecies scattering channel are shifted to $a_\text{r}=0^-$, while the resonance position of the interspecies channel is unshifted still at $1/a_\text{r}=0$ \cite{BCSOC2}.  From Eq.~\eqref{intra-}, we can obtain the binding energy of intraspecies bound-states given by (see Appendix B for the derivation)
\begin{equation}\label{RSOC}
		\frac{1}{\lambda a}=\sqrt{\widetilde{E}_b+1} + \frac{1}{4} \ln{\frac{\sqrt{\widetilde{E}_b+1}-1}{\sqrt{\widetilde{E}_b+1}+1}} ,
\end{equation}
where $\widetilde{E}_b=E_b/(2\epsilon_{\lambda})$ is the dimensionless binding energy.

\begin{figure}[h!]
		\centering
		\includegraphics[width=0.48\textwidth]{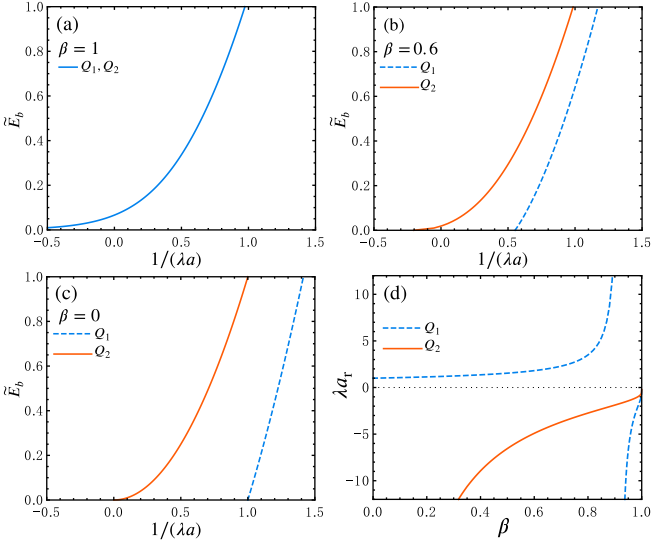}
		\caption{\label{2D SOC}Binding energies and resonance positions of intraspecies bound states of bosons with a 2D SOC, where $\lambda=|\lambda_y|$ and $\beta=\lambda_x^2/\lambda_y^2$. (a) Binding energies in the Rashba SOC case with $\beta=1$. (b) Binding energies with $\beta=0.6$. (c) Binding energies in the 1D case with $\beta=0$. (d) Resonance position $a_\text{r}$ as a function of $\beta$, $a_\text{r}=0^-$ in the Rashba SOC case with $\beta=1$.}
	\end{figure}

In the case of a 2D anisotropic SOC with $\lambda_x \neq \lambda_y$, we can solve binding energies numerically from Eq.~\eqref{intra-} and \eqref{inter-}.  Although the low-energy DOS is qualitatively the same as that without SOC, resonance positions of two intraspecies channels are still shifted due to the quantitative difference, as shown in Fig.~\ref{2D SOC}, while the resonance position of the interspecies channel is unshifted.  In the limit $\beta=\lambda_x^2/\lambda_y^2 \rightarrow 0$, the resonance position of the $Q_2$ scattering channel is unshifted, but the resonance position of the $Q_1$ channel is shifted to $a_\text{r}=1/|\lambda_y|$ due to the anisotropy in Eq.~\eqref{intra-}.  The resonance positions $a_\text{r}$ changing as functions of the aspect ratio $\beta$ is shown in Fig.~\ref{2D SOC}(d).  The difference response to the change in SOC aspect ratio between intraspecies and interspecies channels offers a useful tool to induce resonance in different spin channels.
	
For other types of 2D SOCs, e.g. $\lambda_x=0$ or $\lambda_y=0$, our method can be applied to solve these cases as well.  In the case of $\lambda_x=0$, we find that binding energies of the $Q_2$ intraspecies and the interspecies bound states can be tuned by the SOC strength anisotropy parameter $\lambda_y/\lambda_z$, and resonance can be induced in these two scattering channels, while the binding energy of the $Q_1$ bound state is unaffected.  In the other case with $\lambda_y=0$, binding energies of the $Q_1$ intraspecies and the interspecies bound states can be tuned by the anisotropy of SOC strength $\lambda_z/\lambda_x$, but that of the $Q_2$ bound state is unaffected.

\textit{3D anisotropic SOC}.
With a 3D SOC, as implied in Eq.~\eqref{intra-} and \eqref{inter-}, resonance can be induced in all three scattering channels.  For Weyl SOC, $\lambda_x=\lambda_y=\lambda_z\ne0$, the binding energies of the two intra-species bound states are equal, given by (see Appendix B for the derivation)
	\begin{equation}\label{WSOC}
		\frac{1}{\lambda a}=\frac{2}{3}\sqrt{\widetilde{E}_b+1}+\frac{1}{3}(\sqrt{\widetilde{E}_b}-1/\sqrt{\widetilde{E}_b}) .
	\end{equation}
The binding energy of the interspecies bound state satisfies the same equation as Eq.~\eqref{WSOC} except that $a$ is replaced by $a'$.  The resonance positions of all three bound states are at scattering lengths $0^-$ due to the special low-energy DOS with Weyl SOC \cite{BCSOC3}.

\begin{figure}[h!]
		\centering
		\includegraphics[width=0.48\textwidth]{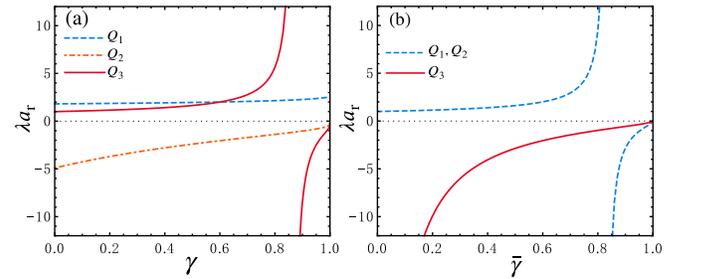}
		\caption{\label{3D_SOC}Resonance positions of bosons with 3D anisotropic SOC where $\lambda=\max(|\lambda_x|,|\lambda_y|,|\lambda_z|)$, (a) with $\lambda_x^2=0.6 \lambda_y^2$, resonance positions of $Q_2$ and $Q_3$ channels go to $0^-$ when $\gamma=\lambda_z^2/\lambda_y^2=1$; (b) with $\lambda_x=\lambda_y$ and $\bar{\gamma}=\lambda_x^2/\lambda_z^2<1$, all the resonance positions are away from $0^-$.}
	\end{figure}

For a general 3D SOC with $\lambda_x,\lambda_y,\lambda_z \neq 0$, resonance positions $a_\text{r}$ can be numerically obtained from Eq.~\eqref{intra-} and \eqref{inter-}.  As shown in Fig.~\ref{3D_SOC}(a),  when $\lambda_x$ and $\lambda_y$ are fixed and $\lambda_z$ varies, the resonance position of the interspecies $Q_3$ scattering channel changes much more rapidly than those of intraspecies channels.  Similarly, the resonance position of the intraspecies $Q_1$ ($Q_2$) channel can be effectively tuned by changing SOC strength $\lambda_x$ ($\lambda_y$).

Also shown in Fig.~\ref{3D_SOC}(a), when two lager SOC strengths $\lambda_y$ and $\lambda_z$ are equal, resonance positions of the $Q_2$ and $Q_3$ channels are at scattering lengths $0^-$, while that of $Q_1$ channel is surprisingly stays at a finite value which can be obtained analytically (see Appendix C for the derivation)
\begin{equation}\label{BEQ}
\frac{1}{\lambda a_\text{r}}=1+\beta\frac{\sqrt{1-\beta}-(1-\beta)-\ln(1+\sqrt{1-\beta})}{(1-\beta)^{3/2}},
\end{equation}
where $\beta=\lambda_x^2/\lambda_y^2$.  The unexpected behavior can not be simply interpreted by the enhanced low-energy DOS which is finite at the threshold.  It is rather due to the direction dependences in r.h.s. of Eq.~\eqref{intra-} and~\eqref{inter-} which may cancel the singular behavior in the DOS.  It again shows that the resonance position in each spin channel can be tuned almost separately.

If two smaller SOC strengths are equal, all the resonance positions are away from $a_\text{r}=0^-$, because the low-energy DOS is qualitatively different from the previous case, as shown in Fig.~\ref{DOS}.  In this case, all the resonance positions varies with the smaller SOC strength as shown in Fig.~\ref{3D_SOC}(b), but that of the scattering channel corresponding to the largest SOC strength varies more slowly.  For $\lambda_x=\lambda_y$ and $\bar{\gamma}=\lambda_x^2/\lambda_z^2<1$, we analytically obtain resonance positions of intraspecies bound states (see Appendix C for the derivation)
\begin{equation}
\begin{aligned}\label{sintra}
\frac{1}{\lambda a_\text{r}}=&1-\frac{\bar{\gamma}}{4}\bigg[\frac{\pi}{\sqrt{1-\bar{\gamma}}}-\frac{2}{1-\bar{\gamma}}+\frac{\pi}{(1-\bar{\gamma})^{3/2}} \\
&-\frac{2\cos^{-1}(\sqrt{1-\bar{\gamma}})}{(1-\bar{\gamma})^{3/2}\sqrt{\bar{\gamma}}}\bigg]
\end{aligned}
\end{equation}
and that of interspecies bound state
\begin{equation}
\begin{aligned}\label{sinter}
\frac{1}{\lambda a_\text{r}}=&1-\frac{1}{2}\bigg[\frac{\pi}{\sqrt{1-\bar{\gamma}}}+\frac{2}{1-\bar{\gamma}}-\frac{\pi}{(1-\bar{\gamma})^{3/2}} \\
&+\frac{2\sqrt{\bar{\gamma}}\cos^{-1}(\sqrt{1-\bar{\gamma}})}{(1-\bar{\gamma})^{3/2}}\bigg],
\end{aligned}
\end{equation}
where $\lambda=|\lambda_z|$.

\textit{Fermion bound state with SOC}.
For a Fermi gas with 1D SOC, resonance can be induced by tuning the intensity and detuning energy of the Raman lasers \cite{socir1,socir2,socir3}.  Here we address this problem in the case of a 2D or 3D anisotropic SOC.  We apply the same method and obtain the same equation for the bound-state eigenenergy as Eq.~\eqref{eigeneq}. However, since there is no s-wave interaction between fermions of the same internal state due to Pauli exclusion principle, the coupling matrix $G$ is now replaced by
	\begin{equation}
		G=
			\begin{bmatrix}
				0{\phantom {0}} & 0{\phantom {0}} & 0 & 0 \\
				0{\phantom {0}} & 0{\phantom {0}} & 0 & 0 \\
				0{\phantom {0}} & 0{\phantom {0}} & g_{\uparrow\!\downarrow} & 0 \\
				0{\phantom {0}} & 0{\phantom {0}} & 0 & g_{\downarrow\!\uparrow}
			\end{bmatrix} .
	\end{equation}
There is only one nontrivial solution satisfying Fermi statistics, i.e. the interspecies bound state with $Q_4=[0,0,q_4,-q_4]$ as mentioned previously.  Its eigenenergy $E_0$ is given by the equation
	\begin{equation}\label{feq}
		\frac{m}{4\pi \hbar^2a^\prime}=\frac{1}{V}\sum_\mathbf{k}\left[\frac{1}{2\epsilon_\mathbf{k}} - \frac{1}{4\varepsilon_{\mathbf{k}}^+ - 2E_0} - \frac{1}{4\varepsilon_{\mathbf{k}}^- -2E_0} \right] ,
	\end{equation}
which can be solved analytically in the case with a Rashba SOC
\begin{equation}
		\frac{1}{\lambda a^\prime} =\sqrt{\widetilde{E}_b+1} + \frac{1}{2} \ln{\frac{\sqrt{\widetilde{E}_b+1}-1}{\sqrt{\widetilde{E}_b+1}+1}},
\end{equation}
and in the case with a Weyl SOC
\begin{equation}
		\frac{1}{\lambda a^\prime} =\sqrt{\widetilde{E}_b}-1/\sqrt{\widetilde{E}_b},
\end{equation}
consistent with Ref. \cite{Ftsoc2,Ftsoc3,Ftsoc4}.  In both cases, binding energies vanish as $a^\prime \rightarrow 0^-$ indicating that two fermions can form a bound state for arbitrary $a^\prime$.

\begin{figure}[h!]
		\centering
		\includegraphics[width=0.3\textwidth]{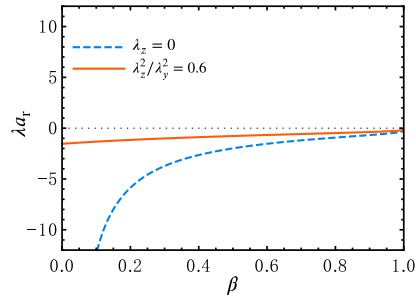}
		\caption{\label{fermionF}Resonance position of fermions with 2D and 3D SOCs vs $\beta=\lambda_x^2/\lambda_y^2$ where $\lambda=|\lambda_y|$.  The solid line is the resonance position for $\lambda_z^2=0.6\lambda_y^2$, and the dashed line is for $\lambda_z=0$.  Both go to $0^-$ when two larger SOC strengths are equal.}
	\end{figure}

For anisotropic SOCs, we solve the binding energy numerically and find that the resonance position is shifted to a finite negative value.  As shown in Fig.~\ref{fermionF}, the resonance position is unshifted in the 1D-SOC limit, and driven to $0^-$ in the case of a Rashba or Weyl SOC, or when two larger SOC strengths equal.  If two smaller SOC strengths are equal, we find the explicit form of resonance position given by
\begin{equation}\label{sf}
\lambda a_\text{r}=-\frac{2}{\pi}\frac{\sqrt{1-\bar{\gamma}}}{\bar{\gamma}},
\end{equation}
where $\lambda_x=\lambda_y$ and $\bar{\gamma}=\lambda_x^2/\lambda_z^2<1$.
As in the boson case, the resonance position can be tuned in cases of 2D and 3D SOCs by changing SOC strengths.

\textit{Discussion and conclusion}.
In experiments, 1D SOC has been generated by coupling Raman lasers to the internal states of atoms \cite{BSOC1}.  In the fermion case resonance position can be tuned near a Feshbach resonance by changing the laser intensity and detuning energy \cite{socir1}.  The 1D case that we have studied in this work corresponds to the limit with zero laser intensity and detuning energy, where SOC induced resonance does not occur.   Mixing of higher partial waves in the effective interaction between dressed atoms was found in the experiment a Bose gas with 1D SOC \cite{socir4}.  We plan to study the boson case with 1D SOC, finite laser intensity and detuning energy in our future works.  A 2D SOC has been generated in a Bose gas in an optical lattice \cite{2SOC1} where on top of the periodic structure the SOC is more complicated than what we have studied.  SOC and confinement induced resonance in this system is an open question.  Near a SOC induced resonance, we expect that the system will suffer severe particle loss, which can be used as an experimental signature to identify the resonance.  The two-body and three-body scattering properties near the SOC induced resonance are important and worth to be studied.  SOC induced resonance can be useful to study macroscopic quantum states \cite{BCSOC2,BCSOC3} and strong correlation effects.

In summary, we find that in an ultracold Bose gas with a general anisotropic 3D SOC, resonances can be induced by tuning SOC strengths.  In a Bose gas with 2D SOC, resonance can be induced in two scattering channels by tunning the aspect ratio of SOC strengths.  In a Bose gas with 3D SOC, resonance in all three scattering channels can be induced.  The resonance position in each scattering channel can be tuned more effectively by each SOC strength in the corresponding direction, which can be very useful for studying spin-dependent effects.  Our method can also be applied to a Fermi gas where resonance can be induced in the case of 2D and 3D SOCs by tuning any SOC strength.

\begin{acknowledgements} We would like to thank Z.-Q. Yu, R. Li, P. Zhang, L. You, and T.-L. Ho for helpful discussions.  This work is supported by NKRDP under Grant No. 2016YFA0301500.
\end{acknowledgements}

\appendix
\section{Derivations of bound-state eigenenergies}
The bound-state eigenenergies generally satisfy Eq.~\eqref{Qeq}
	\begin{equation}{\label{key}}
	Q=\frac{1}{V} G \sum_\mathbf{k} M_{\mathbf{k}}^{-1} Q.
	\end{equation}
By using the SOC inversion symmetry $\mathbf{h}_\mathbf{k}=-\mathbf{h}_{-\mathbf{k}}$, we obtain
\begin{equation}
\sum_\mathbf{k}M_\mathbf{k}^{-1}=
	\begin{bmatrix}
	 A & B & 0 & 0 \\
	B & A & 0 & 0 \\
	0 & 0 & C & D \\
	0 & 0 & D & C \\
	\end{bmatrix} ,
	\end{equation}
where
\begin{align*}
A&=\sum_{\mathbf{k}}\frac{\xi_\mathbf{k}}{\det(M_\mathbf{k})}(\xi_\mathbf{k}^2-2h_{\mathbf{k}z}^2-2h_\mathbf{k}^2), \\
B&=\sum_{\mathbf{k}}\frac{(-2)\xi_\mathbf{k}}{\det(M_\mathbf{k})}(h_{\mathbf{k}x}^2-h_{\mathbf{k}y}^2), \\
C&=\sum_{\mathbf{k}}\frac{\xi_\mathbf{k}}{\det(M_\mathbf{k})}(\xi_\mathbf{k}^2-2h_{\mathbf{k}x}^2-2h_{\mathbf{k}y}^2) , \\
D&=\sum_{\mathbf{k}}\frac{(-2)\xi_\mathbf{k}}{\det(M_\mathbf{k})}(h_{\mathbf{k}x}^2+h_{\mathbf{k}y}^2),
\end{align*}
and $\det (M_\mathbf{k})=\xi_\mathbf{k}^2(\xi_\mathbf{k}^2-4h_\mathbf{k}^2)$.

Therefore we can easily obtain the eigenenergy equation $1/g_{\uparrow\uparrow}=A-B$ for the intraspecies solution $Q_1=(q_1,-q_1,0,0)$, $1/g_{\uparrow\uparrow}=A+B$ for the intraspecies solution $Q_2=(q_2,q_2,0,0)$, and $1/g_{\uparrow\downarrow}=C+D$ for the interspecies solution $Q_3=(0,0,q_3,q_3)$, as in Eq. \eqref{intra-} and \eqref{inter-}.  For the interspecies bound state of fermions with $Q_4=(0,0,q_4,-q_4)$ the eigenenergy equation is given by $1/g_{\uparrow\downarrow}=C-D$, as in Eq. \eqref{feq}.  In these equations, renormalization conditions of coupling constants $m/(4\pi\hbar^2a)=1/g_{\uparrow\uparrow}+\sum_\mathbf{k}1/(2V\epsilon_\mathbf{k})$ and $m/(4\pi\hbar^2a')=1/g_{\uparrow\downarrow}+\sum_\mathbf{k}1/(2V\epsilon_\mathbf{k})$ are used.

\section{Bound-state binding energies with Rashba and Weyl SOC}
For simplicity, we set $\hbar^2/(2m)=1$ in the following derivations. In the case of Rashba SOC, Eq. \eqref{intra-} can be written as
\begin{eqnarray*}
	\frac{1}{8\pi a}&&=\frac{1}{V}\sum_{\mathbf{k}}\left[\frac{1}{2\epsilon_{\mathbf{k}}}-\frac{1}{2\epsilon_{\mathbf{k}}-E_0}\right] \\
&&+\frac{1}{4V}\sum_{\mathbf{k}}\left[\frac{2}{2\epsilon_{\mathbf{k}}-E_0}-\frac{1}{2\varepsilon_{\mathbf{k}}^+-E_0}-\frac{1}{2\varepsilon_{\mathbf{k}}^--E_0}\right],
\end{eqnarray*}
The first summation on r.h.s. of this equation equals to $\dfrac{\lambda}{8\pi}\sqrt{-\widetilde{E}_0}$, where $\widetilde{E}_0=E_0/(2\epsilon_\lambda)$. The second summation can be written as
\begin{eqnarray*}\label{eq-R1}	 &&\frac{1}{4}\frac{1}{(2\pi)^3}\left[\int_{0}^{\infty}d\varepsilon\frac{2D(\epsilon)-D^+(\varepsilon)}{2\varepsilon-E_0}-\int_{-\lambda^2}^\infty d\varepsilon \frac{D^-(\varepsilon)}{2\varepsilon-E_0}\right] \\
&=&\frac{\lambda}{8\pi}\frac{1}{4}\ln\frac{\sqrt{-E_0}-\sqrt{2}\lambda}{\sqrt{-E_0}+\sqrt{2}\lambda},
\end{eqnarray*}
where $D(\varepsilon)=2\pi\sqrt{\varepsilon}\theta(\varepsilon)$ is DOS without SOC, $D^+(\varepsilon)=2\pi(\sqrt{\varepsilon}-\lambda\arctan\sqrt{\varepsilon/\lambda^2})\theta(\varepsilon)$ is DOS of the upper branch with Rashba SOC, and $D^-(\varepsilon)=2\pi[\pi\lambda\theta(\varepsilon+\lambda^2)+(\sqrt{\varepsilon}-\lambda\arctan\sqrt{\varepsilon/\lambda^2})\theta(\varepsilon)]$ is DOS of the lower branch.  It is then straightforward to obtain Eq. \eqref{RSOC}.

In the case of Weyl SOC, Eq. \eqref{intra-} can be written as
\begin{eqnarray*}
\frac{1}{8\pi a}&=&\frac{2}{3V}\sum_\mathbf{k} \left[\frac{1}{2\epsilon_{\mathbf{k}}}+\frac{1}{\xi_{\mathbf{k}}}\right]  \\
 &+&\frac{1}{3V}\sum_\mathbf{k}\left[\frac{1}{2\epsilon_{\mathbf{k}}}+\frac{1}{2}\frac{1}{\xi_{\mathbf{k}}-2h_\mathbf{k}}+\frac{1}{2}\frac{1}{\xi_{\mathbf{k}}+2h_\mathbf{k}}\right],
\end{eqnarray*}
where after the summation the r.h.s. side becomes
$$\frac{\lambda}{8\pi}(\sqrt{\widetilde{E}_b}-1/\sqrt{\widetilde{E}_b}).$$
Eq. \eqref{WSOC} can then be obtained.

\section{Resonance positions with anisotropic SOC}
When two larger SOC strengths are equal, e.g. $\lambda_x^2:\lambda_y^2:\lambda_z^2=1:1:\gamma$, following Eq. \eqref{inter-} the dimensionless binding energy of $Q_3$ bound state is given by
\begin{equation}
	\frac{1}{\lambda a}=\sqrt{\widetilde{E}_b+1}-\frac{4\gamma}{\pi}\int_{0}^{\pi}d\theta\sin{\theta}\cos^2{\theta}I(\theta,\widetilde{E}_b),
\end{equation}
where
\begin{equation*}
I(\theta,\widetilde{E}_b)=\frac{\pi}{8}\frac{-\widetilde{E}_b-1+2\Delta^2+\sqrt{\widetilde{E}_b+1}\sqrt{\widetilde{E}_b+1-\Delta^2}}{\Delta^2\sqrt{\widetilde{E}_b+1-\Delta^2}}
\end{equation*}
and $\Delta=\sqrt{\sin^2\theta+\gamma\cos^2\theta}$.
The integral $\int_{0}^{\pi}d\theta\sin{\theta}\cos^2{\theta}I(\theta,\widetilde{E}_b\to 0)$ is convergent, which leads to a finite resonance position as a function of the anisotropy parameter $\gamma$ given by Eq. \eqref{BEQ}. Meanwhile, the corresponding integral of $Q_1$ or $Q_2$ state, $\int_{0}^{\pi}d\theta\sin^3{\theta}I(\theta,\widetilde{E}_b\to 0)$, is divergent.

Similarly, when two smaller SOC strengths are equal, e.g. $\lambda_x^2:\lambda_y^2:\lambda_z^2=\bar{\gamma}:\bar{\gamma}:1$, the function $\Delta$ becomes $\Delta=\sqrt{\bar{\gamma}\sin^2\theta+\cos^2\theta}$, and the integrand $I(\theta,0)$ becomes
\begin{equation}
	I(\theta,0)=\frac{\pi}{8}\left[\frac{1}{\sqrt{1-\bar{\gamma}}\sin\theta}+\frac{1}{1+\sqrt{1-\bar{\gamma}}\sin\theta}\right]
\end{equation}
The resonance positions are given by
$$\frac{1}{\lambda a_\text{r}}=1-\frac{2\bar{\gamma}}{\pi}\int_0^\pi d\theta\sin^3\theta I(\theta,0)$$
for $Q_{1,2}$ scattering channels as Eq. \eqref{sintra},
$$\frac{1}{\lambda a_\text{r}}=1-\frac{4}{\pi}\int_0^\pi d\theta\sin\theta \cos^2\theta I(\theta,0)$$
for the $Q_3$ channel as Eq. \eqref{sinter}, and
$$\frac{1}{\lambda a_\text{r}}=1-\frac{4}{\pi}\int_0^\pi d\theta\sin\theta\Delta^2 I(\theta,0) $$
for the fermion case as Eq. \eqref{sf}.


\begin{thebibliography}{50}
\bibitem{MFR} C. Chin, R. Grimm, P. Julienne, and E. Tiesinga, Rev. Mod. Phys. \textbf{82}, 1225 (2010).
\bibitem{OFR} P. O. Fedichev, Yu. Kagan, G. V. Shlyapnikov, and J. T. M. Walraven, Phys. Rev. Lett. \textbf{77}, 2913 (1996).
\bibitem{CIR} M. Olshanii, Phys. Rev. Lett. \textbf{81}, 938 (1998).
\bibitem{CIR2} M. Olshanii, Phys. Rev. Lett. \textbf{91}, 163201 (2003).
\bibitem{REV1} J. Dalibard, F. Gerbier, G. Juzeli{\=u}nas, and P. {\"O}hberg, Rev. Mod. Phys. \textbf{83}, 1523 (2011).
\bibitem{REV2} V. Galitski and I. B. Spielman, Nature (London) \textbf{494}, 49 (2013).
\bibitem{REV3} N. Goldman, G. Juzeli{\=u}nas, P. {\"O}hberg, and I. B. Spielman, Rep. Prog. Phys. \textbf{77}, 126401 (2014).
\bibitem{REV4} H. Zhai, Rep. Prog. Phys. \textbf{78}, 026001 (2015).
\bibitem{BSOC1} Y. J. Lin, K. Jim{\'e}nez-Garc{\'\i}a, and I. B. Spielman, Nature \textbf{471}, 83 (2011).
\bibitem{BSOC2} J.-Y. Zhang, S.-C. Ji, Z. Chen, L. Zhang, Z.-D. Du, B. Yan, G.-S. Pan, B. Zhao, Y.-J. Deng, H. Zhai, S. Chen, and J.-W. Pan, Phys. Rev. Lett. \textbf{109}, 115301 (2012).
\bibitem{FSOC1} L. W. Cheuk, A. T. Sommer, Z. Hadzibabic, T. Yefsah, W. S. Bakr, and M. W. Zwierlein, Phys. Rev. Lett. \textbf{109}, 095302 (2012).
\bibitem{FSOC2} P. Wang, Z.-Q. Yu, Z. Fu, J. Miao, L. Huang, S. Chai, H. Zhai, and J. Zhang, Phys. Rev. Lett. \textbf{109}, 095301 (2012).
\bibitem{2SOC1} Z. Wu, L. Zhang, W. Sun, X.-T. Xu, B.-Z. Wang, S.-C. Ji, Y. Deng, S. Chen, X.-J. Liu, and J.-W. Pan, Science \textbf{354}, 83 (2016).
\bibitem{2SOC2} L. Huang, Z. Meng, P. Wang, P. Peng, S.-L. Zhang, L. Chen, D. Li, Q. Zhou, and J. Zhang, Nat. Phys. \textbf{12}, 540 (2016).
\bibitem{socir1} R. A. Williams, M. C. Beeler, L. J. LeBlanc, K. Jim{\'e}nez-Garc{\'\i}a, and I. B. Spielman, Phys. Rev. Lett. \textbf{111}, 095301 (2013).
\bibitem{socir2} Long Zhang, Youjin Deng, and Peng Zhang,  Phys. Rev. A \textbf{87}, 053626 (2013).
\bibitem{socir3} Su-Ju Wang and Chris H. Greene, Phys. Rev. A \textbf{94}, 053635 (2016).
\bibitem{BCSOC2} R. Li and L. Yin, New J. Phys. \textbf{16}, 053013 (2014).
\bibitem{BCSOC3} Dekun Luo and Lan Yin, Phys. Rev. A \textbf{94}, 013609 (2017); \emph{ibid}, Int. J. Mod. Phys. B  \textbf{31}, 1745012 (2017).
\bibitem{Ftsoc2} Zeng-Qiang Yu and Hui Zhai, Phys. Rev. Lett. \textbf{107}, 195305 (2011).
\bibitem{Ftsoc3} Jayantha P. Vyasanakere and Vijay B. Shenoy, Phys. Rev. B \textbf{83}, 094515 (2011).
\bibitem{Ftsoc4} H. Duan, L. You, and B. Gao, Phys. Rev. A \textbf{87}, 052708 (2013).
\bibitem{socir4} R. A. Williams, L. J. LeBlanc, K. Jim{\'e}nez-Garc{\'\i}a, M. C. Beeler, A. R. Perry, W. D. Phillips, I. B. Spielman, Science \textbf{20}, 314 (2012).
\end{thebibliography}
\end{document}